\documentclass[10pt,twocolumn,letterpaper]{article}

\usepackage[pagenumbers]{cvpr} 

\usepackage{graphicx}
\usepackage{amsmath}
\usepackage{amssymb}
\usepackage{booktabs,soul}
\usepackage[pagebackref,breaklinks,colorlinks]{hyperref}

\def\cvprPaperID{25} 
\def\confName{CVPR}
\def\confYear{2022}

\begin{document}

\title{Enhancing VVC with Deep Learning based Multi-Frame Post-Processing}

\author{Duolikun Danier\thanks{Equal contribution.} \qquad\qquad Chen Feng\footnotemark[1] \qquad\qquad Fan Zhang \qquad\qquad David Bull\\
University of Bristol\\
{\tt\small \{duolikun.danier, chen.feng, fan.zhang, dave.bull\}@bristol.ac.uk}
}
\maketitle

\begin{abstract}
  This paper describes a CNN-based multi-frame post-processing approach based on a perceptually-inspired Generative Adversarial Network architecture, CVEGAN. This method has been integrated with the Versatile Video Coding Test Model (VTM) 15.2 to enhance the visual quality of the final reconstructed content. The evaluation results on the CLIC 2022 validation sequences show consistent coding gains over the original VVC VTM at the same bitrates when assessed by PSNR. The integrated codec has been submitted to the Challenge on Learned Image Compression (CLIC) 2022 (video track), and the team name associated with this submission is BVI\_VC.
\end{abstract}

%
\section{Introduction}
\label{sec:intro}

Video compression is one of the most important and popular topics in the image and video processing research field. It plays an essential role to trade off the tension between the large amount of bitrate required for transmitting immersive and high quality video content and the limited bandwidth available \cite{bull2021intelligent}. The efficiency of video codecs have been significantly improved over the past few decades, with the latest MPEG video coding standard, Versatile Video Coding (VVC) \cite{bross2021overview}, achieving nearly 50\% coding gains over its predecessor Higher Efficiency Video Coding (HEVC) \cite{sullivan2012overview}. 

More recently, inspired by the advances of machine learning techniques, in particular with deep convolutional neural networks, a number of deep learning based video coding methods have been proposed. Some of these are designed to offer alternative solutions to the conventional coding framework using auto-encoder type architectures associated with end-to-end optimization \cite{lu2019dvc,balle2016end}, while another group of methods focus on the enhancement of individual coding tools for standard video codecs \cite{zhang2020enhancing,yan2018convolutional}. All these method have demonstrated great potential to outperform conventional hybrid video coding algorithms. On the other hand, we noted that the aim of video compression is to offer optimal visual quality with a given bitrate rather than to minimize the absolute difference between the coded content and its corresponding original. This concept can be integrated with the deep learning based coding methods using a perceptually-inspired loss function for training and optimization \cite{ma2020gan}.

In this paper, a deep learning based multi-frame post processing approach is presented, which has been submitted to the Challenge on Learned Image Compression (CLIC) 2022 (video track). This method is based on a previously developed perceptual-inspired Generative Adversarial Network (GAN) architecture, CVEGAN \cite{ma2020cvegan}. It allows multiple frames (rather than a single frame) as input, which further improves the overall  enhancement performance. This approach has been integrated with the Versatile Video Coding Test Model, VTM 15.2, and it achieves consistent coding gains based on the assessment of PSNR when tested on the CLIC validation video sequences. 

The rest of the paper is organized as follows. Section \ref{sec:method} describes the multi-frame post-processing method, the integrated coding framework and the training process. The coding results are then presented in Section \ref{sec:experiments}. Finally, Section \ref{sec:conclusion} concludes the paper and outlines the future work. 

\section{Proposed Algorithm}
\label{sec:method}

\begin{figure}[t]
\centering
\includegraphics[width=\linewidth]{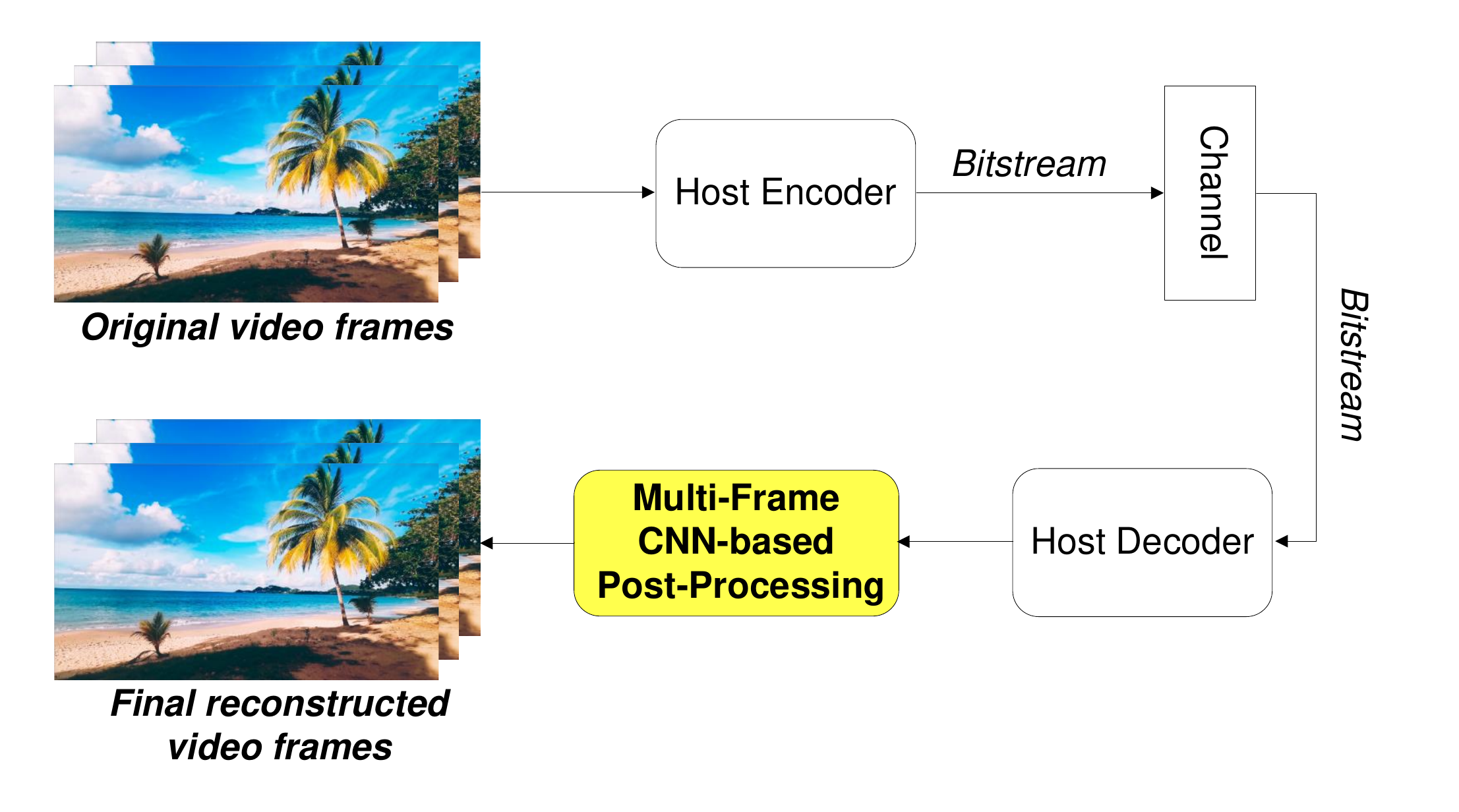}
\vspace{-5mm}
\caption{The coding framework with a CNN-based multi-frame post-processing module.\label{fig:pp}}
\vspace{-5mm}
\end{figure}

The coding framework with the multi-frame post-processing approach is shown in Fig \ref{fig:pp}. The encoder process is identical to that in standard video codecs, and we use VVC VTM 15.2 \cite{s:VVC1} as the host encoder. The CNN-based post-processing is applied at the decoder after the host decoder reconstructs video frames from the compressed bitstream. The employed network architecture for multi-frame post-processing and its training process are described below.

\subsection{Employed Network Architecture}

In this work, we used the same generator architecture of the Generative Adversarial Network for Compressed Video quality Enhancement (CVEGAN), which was originally developed for single frame post-processing and spatial resolution adaptation. CVEGAN has been reported of offer superior coding gains compared to other state-of-the-art network structures when integrated into various coding modules and host codecs \cite{ma2020cvegan}. 

The only difference from the original CVEGAN, where the network processes an input block segmented from a single frame, is that a 96$\times$96$\times$9 input patch is accepted by the generator network in this work. It is obtained by cropping three 96$\times$96$\times$3 YCbCr 4:4:4 blocks from three consecutive reconstructed frames (at the same spatial location), and combining them as a nine channel patch. The network output is in the same format, targeting their uncompressed counterpart.  

The architecture of the discriminator also remains the same as the CVEGAN, which takes the output of the generator and the ground truth patch as the input, and outputs a set of feature points for calculating the discriminator loss. More details on the CVEGAN architecture and its training methodology can be found in \cite{ma2020cvegan}. 

\subsection{Training Configuration}

We also follow the same training strategy as for the original CVEGAN \cite{ma2020cvegan}, which consists of two stages. First, the generator is trained using a combined perceptual loss function to obtain the preliminary model. The used loss function is given as below.
\begin{equation}
\label{equ:equation}
    \mathcal{L}_{p} = 0.3\mathcal{L}_{L1} + 0.2\mathcal{L}_{SSIM} + 0.1\mathcal{L}_{L2}+ 0.4\mathcal{L}_{MSSSIM}
    \end{equation}
The generator is then trained jointly with the discriminator using the ReSphereGAN training methodology\cite{ma2020cvegan}. 

The employed network was implemented based on the PyTorch platform version 1.10 \cite{Paszke2019}. The training process was performed based on the following configurations: Adam optimization \cite{kingma2014adam} with the hyper-parameters: $\beta_1$=0.9 and $\beta_2$=0.999.; batch size of 16; 200 training epochs (100 for both Stage 1 and 2); initial learning rate (0.0001); weight decay of 0.1 for every 100 epochs.

\subsection{Training Content}

The training data was generated from 200 HD source sequences in the BVI-DVC \cite{ma2021bvi} database, and 562 videos clips (with a spatial resolution of 720p) from the YouTube User Generated Content (UGC) dataset \cite{wang2019youtube}. BVI-DVC has been used by MPEG JVET as a training database for optimizing neural network based coding tools of VVC, while YouTube UGC contains diverse content which has similar characteristics to the CLIC validation set. 

All the original sequences were encoded using VVC VTM 15.2 Random Access mode with two quantization parameter (QP) values  (32 and 46). These two QP values were selected to simulate the scenarios for two target bitrates (1 Mbps and 0.1 Mbps) set up by the CLIC 2022. All the compressed sequences and their original counterparts were then cropped into 96$\times$96$\times$9 patches and randomly selected as the training material. Rotation and flips were also used for data augmentation. This results in 80,000 pairs of patches in total. After training, two CNN models are obtained for two bitrate scenarios (1 Mbps and 0.1 Mbps).

\section{Results and Discussion}
\label{sec:experiments}

\begin{figure*}
    \centering
    \begin{minipage}[b]{0.275\linewidth}
\centering
\centerline{\includegraphics[width=\linewidth]{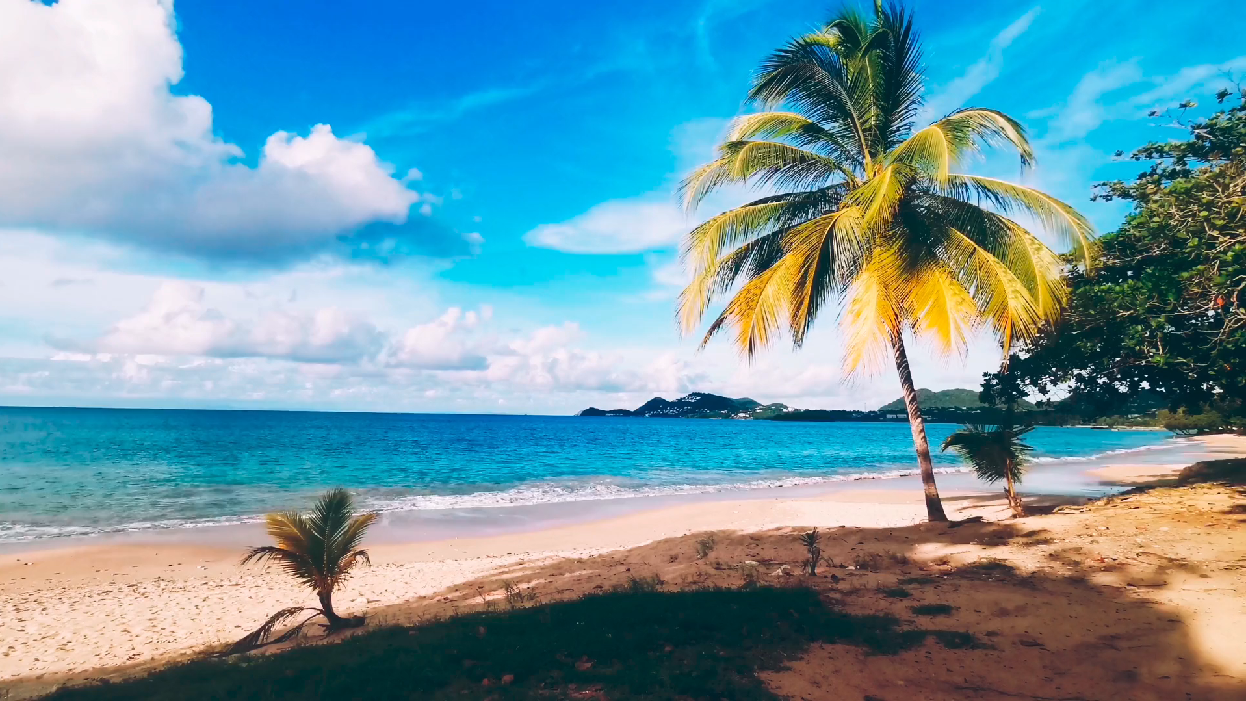}}
S1
\end{minipage}
    \begin{minipage}[b]{0.16\linewidth}
\centering
\centerline{\includegraphics[width=\linewidth]{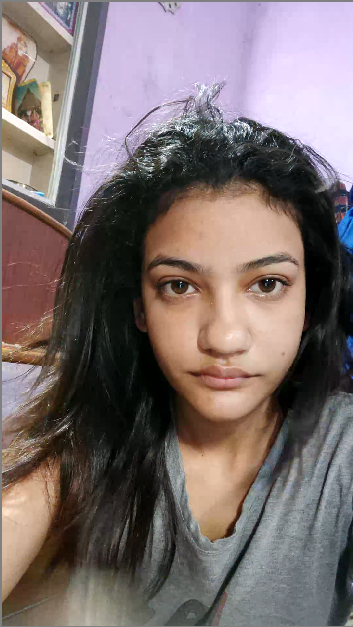}}
S2
\end{minipage}
    \begin{minipage}[b]{0.275\linewidth}
\centering
\centerline{\includegraphics[width=\linewidth]{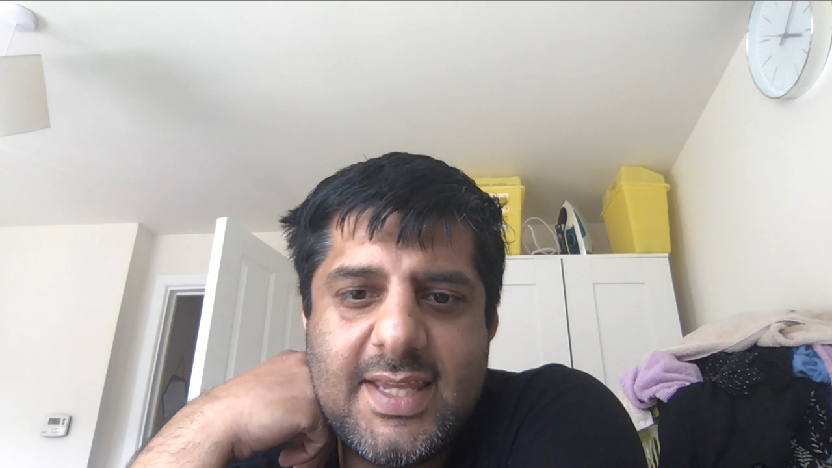}}
S3
\end{minipage}
    \begin{minipage}[b]{0.275\linewidth}
\centering
\centerline{\includegraphics[width=\linewidth]{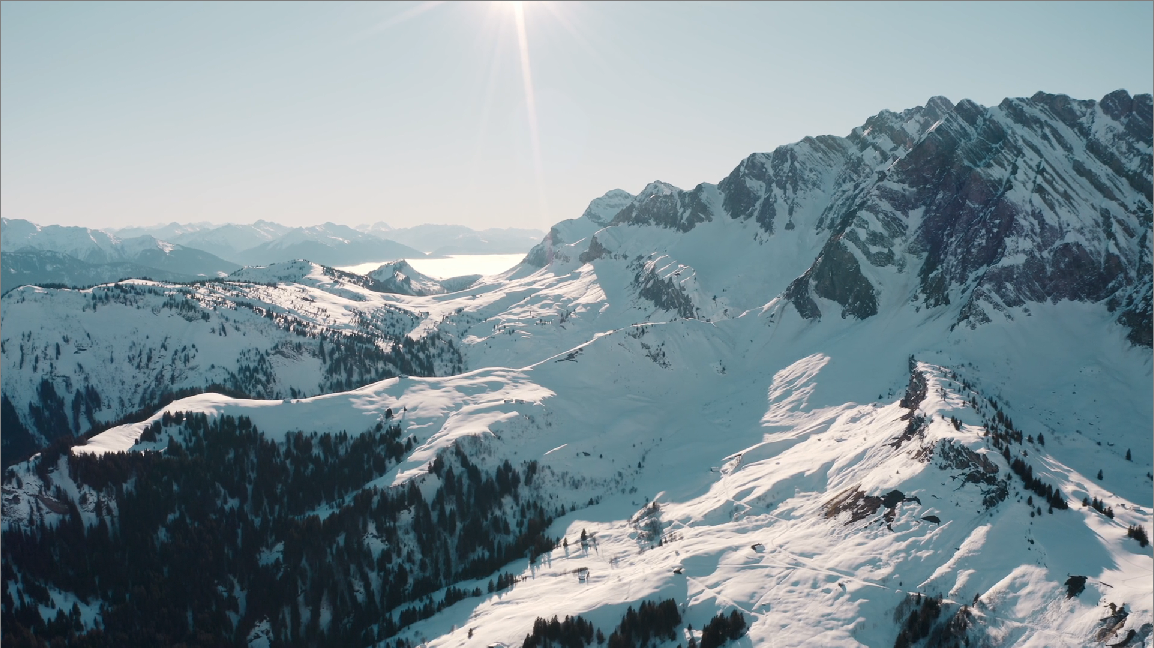}}
S4
\end{minipage}
    \caption{Example frames of the four test sequences.}
    \label{fig:seq}
\end{figure*}

To evaluate the performance of the proposed coding framework, four sequences from the CLIC 2022 validation set was used here for testing the proposed method (the CLIC 2022 test set was not available when the paper was submitted). Their indices and example frames are shown in Figure \ref{fig:seq}. During evaluation, these sequences are first encoded using VVC VTM 15.2 Random Access mode \cite{s:JVETCTC} with a QP value of 46. The bitstreams are then decoded using the VVC VTM decoder and converted to YCbCr 4:4:4 format. Each frame together with its temporally previous and subsequent neighbors are then segmented into 96$\times$96$\times$9 overlapping patches (96$\times$96$\times$3 from each frame at the same spatial location) with a spatial overlap size of 4 pixels as network input. The middle three channels of generator output patch (96$\times$96$\times$3) are then converted to RGB format (required by the CLIC 2022) and aggregated following the same pattern to form the final reconstructed current frame. In the cases when processing the first or the last frame of a sequence, we input 96$\times$96$\times$9 patches cropped from this and two subsequent (or previous) frames, and take the first (or the last) three channels of the generator output to form the final reconstructed frame. The training and evaluation operations were executed on a cluster computer with 32 GPU nodes with 2.4GHz Intel CPUs and NVIDIA P100 GPUs.

The proposed method is benchmarked against the original VVC VTM 15.2, using PSNR for quality assessment. 
Table \ref{tab:results} summarizes the performance of the proposed post-processing method for five different sequences, with an average PSNR gain of 0.09 dB over the original VTM. Here the bitrate remains the same for both codecs (VTM and the proposed method) in each test case. 

\begin{table}[htbp]
    \centering
    \begin{tabular}{c|c cccc}
    \toprule
         Sequence No &  S1 & S2 & S3 & S4\\
         \midrule
         PSNR Gain & 0.08dB & 0.04dB& 0.13dB& 0.1dB\\
         \bottomrule
    \end{tabular}
    \caption{PSNR gains achieved by the proposed method over the original VVC VTM 15.2.}
    \label{tab:results}
\end{table}

\section{Conclusion}
\label{sec:conclusion}

In this paper, we present a CNN-based multi-frame post processing approach for enhancing the visual quality of compression content. This method is based on the perceptual-inspired CVEGAN, and has been integrated with the Versatile Video Coding Test Model (VTM) 15.2 as a submission (BVI\_VC) to the Challenge on Learned Image Compression (CLIC) 2022 (video track). This approach has been evaluated on the CLIC 2022 validate sequences, and the results show consistent coding gains based on the assessment of PSNR. Future work should focus on the complexity reduction of the employed network architecture and more advanced structures for multi-frame processing.

\section*{Acknowledgement}
Duolikun Danier was funded by the China Scholarship Council, University of Bristol, and the UKRI MyWorld Strength in Places Programme (SIPF00006/1).

{\small
\bibliographystyle{ieee_fullname}
\bibliography{egbib}
}

\end{document}